\documentclass[]{article}

\usepackage{amsmath,amssymb}
\usepackage{wasysym}
\usepackage{slashed}
\usepackage{amsfonts}
\usepackage{amssymb}
\usepackage{float}
\usepackage{graphicx,color}
\usepackage{bm}
\usepackage[colorlinks,linkcolor=blue,citecolor=green]{hyperref}

\usepackage{cite}

\usepackage[english]{babel}
\usepackage[latin1]{inputenc}
\usepackage{times}
\usepackage[T1]{fontenc}

\usepackage{amsmath}
\usepackage{bm}
\usepackage{wasysym}
\usepackage{textpos}

\usepackage{graphics}
\usepackage{graphicx}
\usepackage[dvipsnames]{xcolor}
\usepackage{fancybox}
\usepackage{wrapfig}

\usepackage{setspace}
\usepackage{textpos}

\definecolor{Pergamen}{RGB}{235,225,200}
\definecolor{LightGray}{RGB}{235,235,230}
\definecolor{PaleBlue}{RGB}{190,210,255}
\definecolor{DarkGreen}{RGB}{0,80,20}
\definecolor{SoftRed}{RGB}{255,220,170}
\definecolor{DarkBlue}{RGB}{0,10,80}
\definecolor{DarkGray}{RGB}{90,95,95}

\usepackage{authblk}

\title{Non-Extensive Quantum Statistics with Particle -- Hole Symmetry}
\author[1]{T.~S.~Bir\'o\thanks{Biro.Tamas@wigner.mta.hu}}
\author[2]{K.~M.~Shen\thanks{shenkm@mails.ccnu.edu.cn}}
\author[2]{B.~W.~Zhang\thanks{bwzhang@mail.ccnu.edu.cn}}
\affil[1]{HIRG, HAS Wigner Reserach Centre for Physics, Budapest}
\affil[2]{IOPP, Central China Normal University, Wuhan}

\begin{document}
\maketitle


\newcommand{\vs}{\vspace{3mm}}

\newcommand{\be}{\begin{equation}}
\newcommand{\ee}[1]{\label{#1} \end{equation}}
\newcommand{\ba}{\begin{eqnarray}}
\newcommand{\ea}[1]{\label{#1} \end{eqnarray}}
\newcommand{\nl}{\nonumber \\}
\newcommand{\re}[1]{(\ref{#1})}
\newcommand{\spr}[2]{\vec{#1}\cdot\vec{#2}}
\newcommand{\ave}{\overline{u}}
\newcommand{\ve}[1]{\left\vert #1  \right\vert}
\newcommand{\exv}[1]{\left\langle \, {#1} \, \right\rangle}
\newcommand{\inti}{\int_0^{\infty}\limits\!}

\newcommand{\pd}[2]{ \frac{\partial #1}{\partial #2}}
\newcommand{\pt}[2]{ \frac{d #1}{d #2}}
\newcommand{\pv}[2]{ \frac{\delta #1}{\delta #2}}

\newcommand{\grad}{{\vec{\nabla}}}

\newcommand{\ead}[1]{ {\mathrm e}^{#1} }


\begin{abstract}
Based on Tsallis entropy~\cite{Tsallis-1988} and the corresponding deformed exponential function, 
generalized distribution functions for bosons and fermions
have been used since a while~\cite{Teweldeberhan-2003, Silva-2010}. 
However, aiming at a non-extensive quantum statistics further
requirements arise
from the symmetric handling of particles and holes (excitations above and below the Fermi level).
Naive replacements of the exponential function or
''cut and paste'' solutions fail to satisfy this symmetry and to be smooth
at the Fermi level at the same time.
We solve this problem by a general ansatz dividing the deformed exponential
to odd and even terms and demonstrate that how earlier suggestions,
like the $\kappa$- and $q$-exponential behave in this respect.
\end{abstract}



\normalsize







\section{Introduction}

Since Tsallis suggested to use the non-extensive entropy 
formula, $S_T=\frac{1-\sum^W_{i=1}p_i^q}{q-1}$ in 1988~\cite{Tsallis-1988}, 
the corresponding generalized statistical mechanics have been substantially developed
and spread over many fields of 
application~\cite{Tsallis-book,Solar-1998,Self-grav-1965,Self-gravitating-1985,Landsberg-1984,bhs-1987,Vortex-1980,Dark-matter-1993}.
This non-logarithmic relation between entropy and probability is obviously non-additive,
its non-aditivity is comprised in the parameter $q$, differing from one.
Its precise value is determined by the nature of the physical system under consideration. 
It smoothly reconstructs the Boltzmann--Gibbs--Shannon formula at $q=1$.
The application of non-extensive statistical mechanics
becomes mandatory whenever finite size corrections to the thermodynamical
limit are relevant.  Non-extensive systems are those, which behave as final ones
even at large size.

Several applications have been already investigated in a plethora of
physical problems; both on the phenomenological level, by fitting
power-law tailed distributions, and on the mathematical level, seeking
for more and more general construction rules and formulas.
In particular the deformed exponential function, first identified
by obtaining the canonical distribution to the Tsallis entropy, has been
applied to a variety of physical problems.

Quantum statistics erects novel problems to be solved also in this respect.
The naive replacement of the Euler--exponential with another, deformed
exponential function namely can loose the particle--hole symmetry, inherent
in the traditional Fermi distribution above and below the Fermi level.
In many suggestions for the generalized Bose and Fermi distributions 
the Tsallis' $q-$exponential function,  
\begin{equation}
e_q(x) \: := \:  [1+(1-q)x]^{\frac{1}{1-q}}
\label{int-1}
\end{equation}
is used instead of $e^x$ at the corresponding place in the 
formulas~\cite{Teweldeberhan-2003, Silva-2010, Buyukkilic, Bennini, Chen, Teweldeberhan-2005, Conroy, Conroy2, CleymansWorku1, CleymansWorku2}.

\begin{figure}[H]
\vskip0.04\linewidth
\centerline{
\includegraphics[width = 0.7\linewidth]{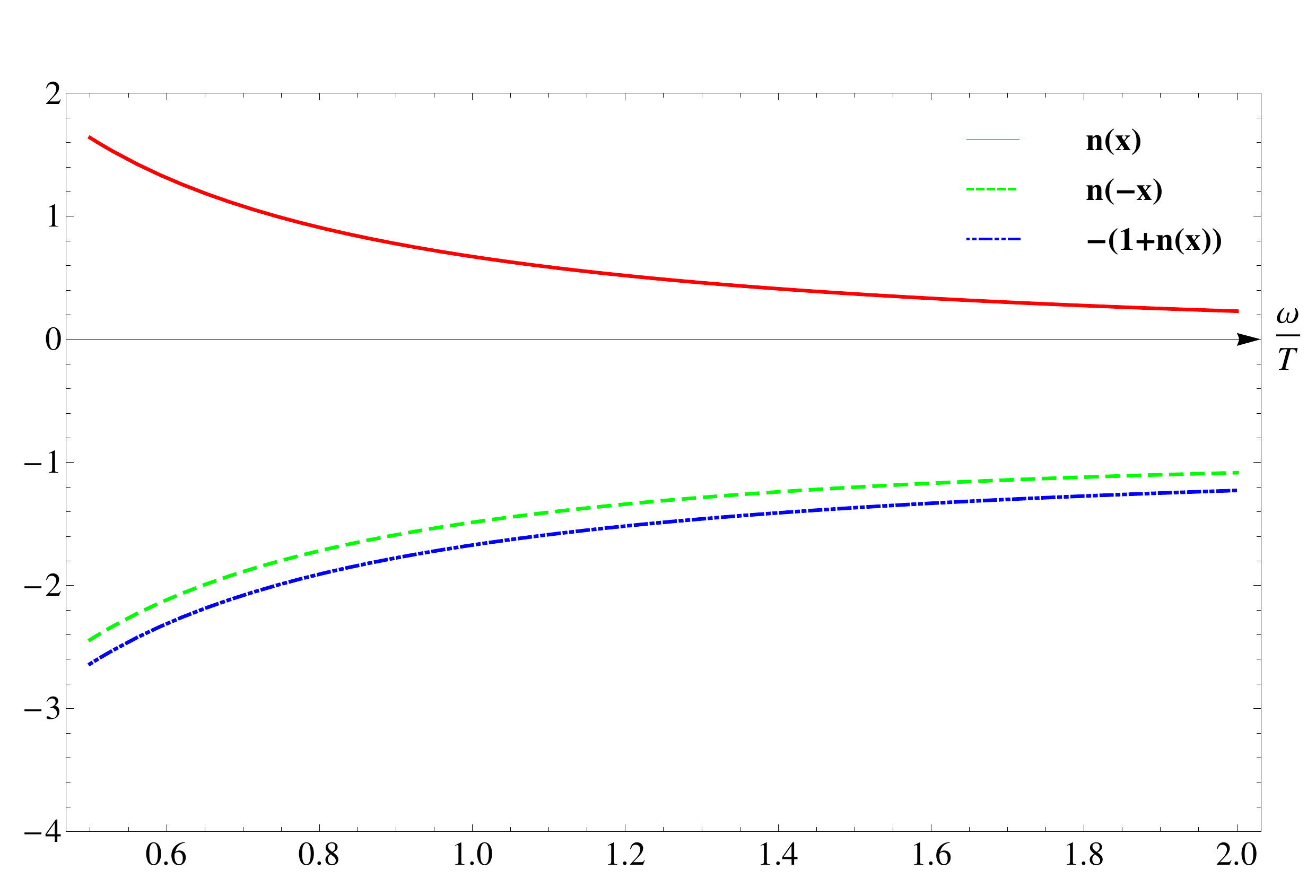}
}
\caption{  \label{int-f1}
The generalized KMS relation for bosons,
tested by using the original Tsallis' $q-$exponential function.
The relation $n(-x)=-1-n(x)$  breaks, since $e_q(x) \ne e_{2-q}(x)=1/e_q(-x)$  
appear in the respective formulas ($q =0.8$).
}
\end{figure}

However, there is a fundamental problem with this ansatz: 
it does not satisfy the CPT invariant concept interpreting holes
among the negative energy states as anti-particles with the corresponding positive 
energy~\cite{Feynman},
\begin{equation}
n(-x)=\mp 1-n(x)
\label{int-2}
\end{equation}
with $x=\omega/T$.  
Here the upper sign is for bosons and the lower one for fermions, respectively. 
At a finite chemical potential (Fermi energy) one uses the argument $x=(\omega-\mu)/T$, and
the above relation expresses a reflection symmetry to the $x=0$ ($\omega=\mu$) case. 
In particular the original $q-$exponential, forming the Tsallis-Pareto cut power-law,
is an incomplete approach in this respect, as long as $q \ne 2-q$.

In this article we explore the general requirement on the deformed exponential 
function used in quantum statistics for satisfying the above symmetry. 
Starting by a generalized form of the Kubo-Martin-Schwinger (KMS) 
relation~\cite{KMS-relation,KMS-relation2}
we derive the desired property that a deformed exponential must satisfy.
Based on this we formulate a suggestion how to "ph-symmetrize" an arbitrary function, 
$e_k(x)$. 


\section{Kubo-Martin-Schwinger Relation}

The KMS relation in its original form states that certain 
correlations between time dependent operators can be related to 
the reversed correlation at finite temperature by shifting the 
time difference variable, $t$, with a pure imaginary shift, $i\beta$:
\begin{flalign}
~~~~~~~~~~~~~~~~~~~~~~~~~~~~~~~~~~~~\langle A_tB_0\rangle
&=\mathrm{Tr}\left(e^{-\beta H}e^{itH}Ae^{-itH}B\right)& \nonumber\\
&=\mathrm{Tr}\left(e^{-\beta H}e^{itH}Ae^{-itH}e^{\beta H}e^{-\beta H}B\right)& \nonumber\\
&=\mathrm{Tr}\left(e^{i(t+i\beta)H}Ae^{-i(t+i\beta)H}e^{-\beta H}B\right)& \nonumber\\
&=\mathrm{Tr}\left(e^{-\beta H}Be^{i(t+i\beta)H}Ae^{-i(t+i\beta)H}\right)& \nonumber\\
&=\langle B_0A_{t+i\beta}\rangle
\label{kms-1}
\end{flalign}
Considering generalized  thermodynamical formulas we have to reconsider 
the KMS relation in a more general setting. 
Since this relation is proven simply by re-shuffling of operators 
under a trace, it holds very generally:
\begin{flalign}
~~~~~~~~~~~~~~~~~~~~~~~~~~~~~~~~~~~\langle A_t \, B_0\rangle
&=\mathrm{Tr} \left(\rho \: U\!AU^{-1} \, B\right)& \nonumber\\
&=\mathrm{Tr} \left(\rho \: U\!AU^{-1} \, \rho^{-1}\!\rho \, B\right)& \nonumber\\
&=\mathrm{Tr} \left(\rho \, B \: (\rho U)A(\rho U)^{-1} \, \right)& \nonumber\\
&=\langle B_0 \, A_{t\oplus i\beta}\rangle
\label{kms-2}
\end{flalign}
We note that for non-exponential $H-$dependence of the density matrix $\rho$ 
the $\oplus$ operation is energy dependent. 
If we accept the requirement that $\rho(H)$ is the analytic continuation
of a unitary function of $H$, then for a general $\rho=g(iH)$, 
from $W^{-1}=W^{\dag}$ one derives $g(iH)^{-1}=g^*(iH)=g(-iH)$ and consequently
\begin{equation}
\frac{1}{\rho(\omega)}=\rho^*(\omega)=\rho(-\omega).
\label{kms-3}
\end{equation}
On the other hand, applying the original KMS relation, Eq.(\ref{kms-1}) for $A_t=e^{-i\omega t}a$ 
and $B_t=A_t^\dag=e^{i\omega t}a^\dag$ one obtains the known relation 
between the occupation density of negative energy states (holes in the positive energy
continuum) and positive energy anti-particles
\begin{eqnarray}
\langle aa^\dag\rangle=\langle a^\dag a\rangle \, \ead{\beta\omega}.
\label{kms-4}
\end{eqnarray}
With the commutation and anti-commutation relation of basic operators
\begin{eqnarray}
[a,a^\dag]_{\mp}=1
\label{kms-5}
\end{eqnarray}
we arrive at the well-known distribution functions for the occupation density
\begin{eqnarray}
	n(\beta\omega)= \exv{a^{\dag}a} = \frac{1}{e^{\beta\omega}\mp 1},
\label{kms-6}
\end{eqnarray}
where the upper sign is for bosons and the lower one for fermions, respectively.

As a consequence a missing negative energy boson state is equivalent with a positive energy 
bosonic hole state: $-n(-\omega)=(1+n(\omega)) >0$. 
The connection to the canonical thermodynamical weight is given by
\begin{equation}
f(\omega)=\frac{n(\omega)}{1+n(\omega)}=e^{-\beta\omega}.
\label{kms-7}
\end{equation}
Here the exponential factor is to be generalized 
$e^{-\beta\omega} \rightarrow e_k(-\beta\omega)$. 
However, an arbitrary guess of a function for it will not satisfy the 
generalized KMS relation: using a deformed exponential $e_k(x)$
in Eq.(\ref{kms-7}) the relation
\begin{equation}
e_k(x)\cdot e_k(-x):=\frac{n(\omega)}{1+n(\omega)}\frac{n(-\omega)}{1+n(-\omega)}=1
\label{kms-8}
\end{equation}
follows.
The originally suggested cut power-law, the $q$-exponential in Eq.(\ref{int-1}) \cite{Tsallis-1988},
\begin{equation}
e_q(x)\equiv (1+k x)^{1/k}
\label{kms-9}
\end{equation}
with $k=1-q$, although obtained by physical arguments and explored in manifold experimental data, 
does not directly satisfy this relation.
On the other hand, the $\kappa-$exponential, suggested by Kaniadakis~\cite{Kaniadakis-2001},
\begin{equation}
e_{\kappa}(x)\equiv (\sqrt{1+(\kappa x)^2}+\kappa x)^{1/\kappa}
\label{kms-10}
\end{equation}
does, as it can be proven by direct substitution.


\section{General Particle-Hole Symmetry within Nonextensive Quantum Statistics}

In the followings we study the general form of deformed exponential functions satisfying the 
relation\footnote{We note that using the formal logarithm, due to $e_k(x)=\ead{L(x)}$
this requirement is simply the oddity of $L(x)=-L(-x)$.}
\begin{equation}
e_k(x)\cdot e_k(-x)=1.
\label{phs-1}
\end{equation}
The most general real ansatz with power-law asymptotics $x^{1/k}$ is given 
as
\begin{equation}
e_k(x):=(a+bk x)^{1/k}
\label{phs-2}
\end{equation}
with $a(x^2)$ and $b(x^2)$ being even functions of the variable $x$. The power parametrization by $k$
ensures the inclusion of the Boltzmann-Gibbs limit by
\begin{equation}
\lim_{k\rightarrow 0}e_k(x)=e^x,
\label{phs-3}
\end{equation}
if
\begin{equation}
\lim_{k\rightarrow 0}a(x^2)=\lim_{k\rightarrow 0}b(x^2)=1.
\label{phs-4}
\end{equation}
Utilizing now the requirement of Eq.(\ref{phs-1}), one obtains
\begin{equation}
e_k(x)\cdot e_k(-x)=(a^2-b^2k^2 x^2)^{1/k}=1,
\label{phs-5}
\end{equation}
leading to the general form
\begin{equation}
e_k(x)=(\sqrt{1+(bk x)^2}+bk x)^{1/k}.
\label{phs-6}
\end{equation}
It is straightforward to realize that the simplest choice, $b=1$ and $k=\kappa$, 
leads to the Kaniadakis $\kappa$-exponential~\cite{Kaniadakis-2001} cf. Eq.(\ref{kms-10}).

There are, however, other solutions satisfying the particle-hole symmetry requirement:
\begin{equation}
e_k(x)=\left(\frac{f(-x)}{f(x)}\right)^{1/k} 
= \left(\frac{1+\lambda ~k x}{1-\lambda ~k x}\right)^{1/k}
\label{phs-7}
\end{equation}
with $\lambda(x^2)$ being an even function of $x$. 
This form is equivalent with the $k-$form above Eq.(\ref{phs-6}), revealing the connection
\begin{equation}
b=\frac{2\lambda}{1-\lambda^2(k x)^2}.
\label{phs-8}
\end{equation}
With respect to the Tsallis' $q-$exponential ($q=1-k$), that ansatz may be 
slightly modified to a ratio, according to Eq.(\ref{phs-7}) by putting $\lambda=1/2$,
\begin{equation}
\tilde{e}_q(x):=e_q(x/2)\cdot e_{q^*}(x/2)=\left(\frac{1+(1-q)x/2}{1-(1-q)x/2}\right)^{\frac{1}{1-q}}.
\label{phs-9}
\end{equation}
This ansatz is equivalent to the general formula Eq.(\ref{phs-6}) with $1/b=1-k^2 x^2/4$. 
Based on this the corresponding Bose distribution function becomes
\begin{equation}
\tilde{n}(x)=\frac{1}{\tilde{e}_q(x) - 1}=\frac{\left[1-(1-q)\frac{x}{2}\right]^{\frac{1}{1-q}}}
{\left[1+(1-q)\frac{x}{2}\right]^{\frac{1}{1-q}} - \left[1-(1-q)\frac{x}{2}\right]^{\frac{1}{1-q}}}.
\label{phs-10}
\end{equation}
Another parametrization,
which reflects the relation between the $\kappa-$ and symmetrized $q-$exponential, 
is given by~\cite{KanEPJA40}
\begin{equation}
	\left[e_{k}(x)\right]^k = e^z, 
	\quad bk x=\sinh z, 
	\quad \lambda k x=\tanh\frac{z}{2}.
\label{phs-11}
\end{equation}
This leads us to a general procedure how to ph-symmetrize a 
suggested deformed exponential function with power-law asymptotics, $e_q(x)$:
\begin{enumerate}
\item From the originally suggested base of the asymptotic power we compose 
	the $b(x^2)$ function according to
\begin{equation}
b=\frac{[e_q(x)]^k-[e_q(-x)]^k}{2k x}.
\label{phs-12}
\end{equation}

\item Using Eq.(\ref{phs-6}) the ph-symmetrized deformed exponential becomes
\begin{equation}
e_{k}(x)=(\sqrt{1+(bk x)^2}+bk x)^{1/k}.
\label{phs-13}
\end{equation}

\item This function, $e_{k}(x)$, should be used in the formulas for Bose and Fermi distributions.
\end{enumerate}
We note that starting with the original Tsallis-Pareto form, Eq.(\ref{kms-9}), after the above steps
$b=1$ and one exactly obtains the Kaniadakis form, Eq.(\ref{kms-10}).

There is another way to construct ph-symmetric Bose distributions, namely one may search for a linear
combination of traditional $n(\omega)$ formulas with the suggested 
$e_q(x)$ on the one hand and its dual, $e_{q^*}(x)=1/e_q(-x)$ on the other hand. Such an ansatz,
\begin{equation}
n_{\mathrm{KMS}}(x)=A(n_q(x)+n_{q^*}(x))+B
\label{phs-14}
\end{equation}
for the Tsallis distribution with $q^*=2-q$, finally leads to $B=A-\frac{1}{2}$ and for having
"zero for zero" we arrive at
\begin{equation}
n_{\mathrm{KMS}}(x)=\frac{1}{2}\left(n_q(x)+n_{q^*}(x)\right).
\label{phs-15}
\end{equation}
It is not trivial, which $b(x^2)$ function corresponds to this choice. 
Although the corresponding weight factor obviously satisfies the relation of Eq.(\ref{kms-8}),
it is easy to see that in this procedure
always one of the parameters $q$ and $q^*$ is less than one, cutting off
the asymptotically high $x$-tail of the distribution.

Summarizing this part, a single even function, $b(x^2)$ determines the 
quantum statistical ansatz with a proper ph-symmetry
\begin{equation}
n_{B,F}(x)=\frac{1}{(\sqrt{1+b^2(k x)^2}+b k \, x)^{1/k}\mp 1}.
\label{phs-20}
\end{equation}
This expression has the asymptotics
\begin{equation}
\lim_{x\rightarrow \infty}n_{B,F}(x)=(2bk \, x)^{-1/k},
\label{phs-21}
\end{equation}
provided that $bk$ is positive. On the other hand the equivalent $\lambda(x^2)$ view leads to
\begin{equation}
n_{B,F}(x)=\frac{(1-\lambda k x)^{1/k}}{(1+\lambda k x)^{1/k}\mp(1-\lambda k x)^{1/k}}
\label{phs-22}
\end{equation}
reaching $n(x)=0$ at a finite value of the argument. It is defined only for $\lambda k x\le 1$.


Finally we analyze the ''cut and paste'' solution, using alternatingly
the $e_q(x)$ and $e_{q^*}(x)$ dual deformed exponentials depending
on the sign of the argument, as it has been introduced by
A. M. Teweldeberhan \textit{et al.}~\cite{Teweldeberhan-2005}:
\be
n_F(x) = \left\{ \begin{array}{c}  
		\frac{1}{e_q(x) + 1} \qquad  x>0 \\ 
		\\	
		\frac{1}{e_{q^*}(x) + 1} \qquad  x<0 	
\end{array} \right.
\ee{CUTANDPASTE}
The requirement $n_F(x)+n_F(-x)=1$ delivers $e_q(x)\cdot e_{q^*}(-x)=1$,
which is satisfied by the original Tsallis ansatz,
\be
 e_q(x) = \left( 1 + k \, x \right)^{1/k},
\ee{ORIGTSA}
with $q=1+k$ and $q^*=1-k$. In this case always
$q+q^*=2$.

In general the deformed Fermi distribution can be expanded around the
Fermi surface ($x=0$) and the even and odd terms can be collected
separately. Ans\"atze based on a single function for all real $x$
values must have the form
\be
n_F(x) = \frac{1}{2} \left(m(x^2) + x \cdot n(x^2) \right).
\ee{EVENODDFERMI}
As a consequence of the KMS relation $n_F(x)+n_F(-x)=m(x^2)=1$ has to be satsified,
and $n_F(x)-n_F(-x)=x \cdot n(x^2)$ is a purely odd function, prohibiting
all even order derivatives at $x=0$.

The cut and paste solutions,
\be
n_F(x) = \left\{ \begin{array}{c}  
		\frac{1}{2} \left(m(x^2) + x \, n(x^2) \right) \qquad  x>0 \\ 
		\\	
		\frac{1}{2} \left(m^*(x^2) + x \, n^*(x^2) \right) \qquad  x<0 \\ 
\end{array} \right. ,
\ee{CAPFERMI}
as a consequence of $n_F(x)+n_F(-x)=1$ on the other hand have to comply with
\ba
 m(x^2) + m^*(x^2) &=& 2,
\nl
 n(x^2) - n^*(x^2) &=& 0.
\ea{COMPLY}
This comprises the deformed Fermi distribution into the following form:
\be
n_F(x) = \left\{ \begin{array}{c}  
	\frac{1}{2} \left(m(x^2) + x \, n(x^2) \right) \qquad  x>0 \\ 
		\\	
	 1 - \frac{1}{2} \left(m(x^2) - x \, n(x^2) \right) \qquad  x<0 \\ 
\end{array} \right. .
\ee{CUT_FERMI}
It is easy to realize that the odd part is given as the half of
\be
n_F(x) - n_F(-x) = \left( m(x^2)-1 \right) \cdot {\mathrm{sign}}(x) \, + \, x \, n(x^2).
\ee{ODD_FERMI}
For not having a jump in the value of $n_F(x)$ at $x=0$ only $m(0)=1$
is necessary, but for being smooth up to arbitrary order in derivatives
at $x=0$ the functional identity $m(x^2)=1$ must be satisfied.
In this case the the deformed Fermi distribution has the general expression
\be
n_F(x) = \frac{1}{2}  \left( 1 \: + \: x \, n(x^2)  \right).
\ee{GEN_DEF_FERMI}
Consequently the expansion around the Fermi surface ($x=0$) contains 
odd terms only. The well-known property of the Sommerfeld expansion~\cite{jump},
an expansion of integrals of a test function multiplied by the original Fermi distribution,
reflects exactly this property: only odd derivatives of the test function
occur in the result.
For the Bose distribution an analogous argumentation
holds, but there at $x=0$ the Bose condensation occurs, the distribution
diverges and therefore finite jumps in even derivatives are only of
theoretical importance.








\section{Summary and Conclusions}

Considering the generalized KMS relation we have established that
quantum statistical distributions using a deformed exponential
function must satisfy the relation (\ref{phs-1}),
$e_k(x)\cdot e_k(-x)=1$, for 
reflecting the particle-hole  symmetry smoothly at the Fermi level.
The use of the original Tsallis' $q-$exponential confronts 
with this requirement, since $e_q(-x)=1/e_{2-q}(x)\ne 1/e_q(x)$,
while the kappa-exponential, promoted by Kaniadakis, is in accord
with this.

We have derived the general formula for deformed exponentials satisfying
this basic requirement, and found that it has a few equivalent forms,
each determined by a single even function. These functions, either $b(x^2)$ or 
$\lambda(x^2)$, are connected in a particular way.
Using this connection, reflecting a general splitting to even and odd terms
of a function, we have pointed out that the $\kappa$-distribution
is the properly ph-symmetric improved pendant of the $q$-distribution.

In this context, some other solutions to this basic requirement
are also mentioned, in particular an arithmetic mean of
two Tsallis-Bose functions with dual $q$, $q^*=2-q$ parameters. 
In general we found that the expansion of the generalized Fermi distribution
with proper symmetry around the Fermi surface contains only odd terms in the argument,
$x=(\omega-\mu)/T$, besides the trivial zeroth order term, $1/2$.
Cut and paste solutions contain a jump starting with the second derivative
at the Fermi level due to their not respecting the above rule.

While this paper concentrated on the analysis of mathematical properties
of generalized quantum statistical particle number distributions,
there should be ample room for physical applications, 
whose discussion - however - has to be delegated to other works.


\section{Acknowledgement}

This work has been supported in part by MOST of China under 2014DFG02050,
the Hungarian National Research Fund OTKA (K104260), a bilateral governmental
Chinese-Hungarian agreement NIH TET$\_$12$\_$CN-1-2012-0016 and
by NSFC of China with Project Nos. 11322546, 11435004.

\end{document}